 \documentstyle[12pt]{article}
 \oddsidemargin--1mm
\begin{document}
 
\begin{center}
 {\LARGE Path Integral Quantization and Riemannian-Symplectic Manifolds}

 \vskip 0.5cm
 Sergei V. SHABANOV $^{a,}$\footnote{On leave from Laboratory 
 of Theoretical
 Physics, JINR, Dubna, Russia.}\ \ \ and\ \ \ John R. KLAUDER$ ^{b}$

 \vskip 0.5cm
 ${}^a${\em Institute for Theoretical Physics, FU-Berlin, Arnimallee 14,
 Berlin,\\ D-14195, Germany }\\
 ${}^b${\em Departments of Physics and Mathematics,
 University of Florida,
 Gainesville, \\ FL-32611, USA}
 \end{center}

\def\R{\relax{\rm I\kern-.18em R}}
\def\1{\relax{\rm 1\kern-.27em I}}
\newcommand{\Z}{Z\!\!\! Z}
\def\<{\langle}
\def\>{\rangle}
\newcommand{\oo}{\stackrel{\circ}{\omega}{}}
\newcommand{\pll}{\partial}

\begin{abstract}
We develop a  mathematically well-defined 
path integral formalism for general symplectic manifolds.
We argue that in order to make
a path integral quantization covariant under general
coordinate transformations on the phase space and involve a 
genuine functional measure that is both finite 
and countably additive, the phase space manifold should be
equipped with a Riemannian structure (metric). A suitable method
to calculate the metric is also proposed. 
\end{abstract}

{\bf 1}. 
Most quantization formulations lead to correct results only when the
system in question is expressed in one of a family of special 
(Cartesian) coordinate
systems \cite{dir2}. However, for Hamiltonian systems on a generic
symplectic manifold, the canonical variables do not admit such
special coordinates. Moreover classical Hamiltonian dynamics
exhibits covariance under general coordinate transformations
on the symplectic manifold, and, in this sense, is coordinate-free.
This covariance is lost upon a formal phase-space path 
integral quantization. The reason is that the Hamiltonian 
action involves terms linear in the time derivatives and, therefore, its
exponential (even in Euclidean time) does not induce a 
proper (normalizable, $\sigma$-additive) measure on the path
space which is covariant under general coordinate transformations.
Since the discovery of the Hamiltonian path integral by Feynman,
the indisputable fact of its coordinate dependence has generally
complicated any straightforward use of Hamiltonian
path integrals in applications. Indeed, to justify any formal manipulation
regarding the path integral measure, the operator formalism has generally
been invoked. 
The aim of this letter is to resolve this long-standing problem and
to establish a rigorous, coordinate-free path integral 
formalism for general Hamiltonian systems. 

In \cite{kl88} it was shown that the formal phase-space path
integral measure can be made covariant under general coordinate
transformations by means of a special regularization that involves
an auxiliary Brownian motion on a flat phase space. The desired
quantum mechanics is restored by taking the diffusion constant
to infinity. The modified 
Hamiltonian path integral determines a stochastic process 
on the phase space in which the original first-order Lagrangian
plays the role of the external potential.  The stochastic process on 
a {\em flat} phase space
remains flat under coordinate transformations. Thus one may change the
variables in the regularized path integral according to the rules
of the stochastic integral calculus, and at the end of calculations
one has to take the infinite diffusion constant limit.

To be specific, and also to introduce notation, consider quantum
mechanics in a flat phase space. 
Let $\hat{q}^j$ be a set of Cartesian coordinate
canonical operators satisfying the
Heisenberg commutation relations $[\hat{q}^j,\hat{q}^k] = i\oo^{jk}$.
Here $\oo^{jk} = -\oo^{kj}$ is the canonical symplectic structure.
We introduce the canonical coherent states as $
|q\> \equiv e^{iq^j\oo_{jk}\hat{q}^k}|0\>$, where
$\oo_{jn}\oo^{nk}=\delta^k_j$,
and $|0\>$ is the ground state of a harmonic oscillator with unit angular
frequency. Any state $|\psi\>$ is given as a function on phase space in this
representation by $\<q|\psi\> =\psi(q)$. A general operator $\hat A$
can be represented in the form $\hat A =\int dq \,a(q)|q\>\<q|$, where
$a(q)$ is the lower symbol of the operator and $dq$ is a properly
normalized form of the Liouville measure. The function $A(q,q') = 
\<q|\hat{A}|q'\>$ is the kernel of the operator. 

The main object of the path integral formalism is the integral kernel 
of the evolution operator
\begin{equation}
K_t(q,q') = \<q|e^{-it\hat H}|q'\> =
\int\limits_{q(0)=q'}^{q(t)=q} {\cal D}q
e^{i\int_0^t d\tau\left(\frac 12 q^j\oo_{jk}\dot{q}^k -h  \right) }\ .
\label{eopk}
\end{equation}
Here $\hat H$ is the Hamiltonian, and $h(q)$ its symbol. The measure
formally implies a sum over all phase-space paths pinned at the initial and
final points,  and a Wiener measure regularization implies the 
following replacement 
\begin{equation}
{\cal D}q\rightarrow {\cal D}\mu_\nu(q)=
{\cal D}q \, e^{-\frac {1}{2\nu}\int_0^t d\tau\, \dot{q}^2}=
N_\nu(t)\,d\mu_W^\nu(q)\ .
\label{wm}
\end{equation}
The factor $N_\nu(t)$ equals $2\pi e^{\nu t/2}$ for every degree of
freedom, $d\mu_W^\nu(q)$ stands for the Wiener measure, 
and $\nu$ denotes the diffusion constant. We denote by $K^\nu_t(q,q')$ 
the integral kernel of the evolution operator for a finite $\nu$.
The Wiener measure determines a stochastic process on the {\em flat}
phase space. The integral of the symplectic one-form $\int q\oo dq$
is a stochastic integral that is interpreted in the Stratonovich
sense. Under general coordinate transformations $q = q(\bar{q})$,
the Wiener measure describes {\em the same} stochastic process
on {\em flat} space in the curvilinear coordinates $dq^2 =
d\sigma(\bar{q})^2$, so that the value of the integral is
not changed apart from a possible phase term. After the
calculation of the integral, the
evolution operator kernel is obtained by taking the limit
$\nu\rightarrow\infty$. The existence of this limit,
 and also the covariance under
general phase-space coordinate transformations, can be most easily
proved through the {\em operator} formalism for the regularized
kernel $K_t^\nu(q,q')$.

Note that the integral (\ref{eopk}) with 
the Wiener measure inserted can be regarded as an ordinary
Lagrangian path integral with a complex action, where the configuration
space is the original phase space and the Hamiltonian $h(q)$ serves as
a potential. Making use of this observation it is not hard to
derive the corresponding Schr\"odinger-like equation
\begin{equation}
\pll_t K_t^\nu(q,q') = \left[\frac{\nu}{2}\left(
\pll_{q^j} + \frac i2 \oo_{jk}q^k\right)^2 - ih(q)\right]K_t^\nu(q,q')\ ,
\label{ofwm}
\end{equation}  
subject to the initial condition
$K_{t=0}^\nu(q,q') = \delta(q-q'), 0<\nu<\infty$.
One can easily show that $\hat{K}^\nu_t \rightarrow
\hat{K}_t$ as $\nu\rightarrow \infty$ for all $t>0$.
The covariance under general coordinate transformations follows
from the covariance of the ``kinetic'' energy of the Schr\"odinger
operator in (\ref{ofwm}): The Laplace operator is replaced by
the Laplace-Beltrami operator in the new curvilinear
coordinates $q = q(\bar{q})$, so the solution is not changed, but
written in the new coordinates. This is similar to the covariance
of the ordinary Schr\"odinger equation and the corresponding
{\em Lagrangian} path integral relative to general coordinate
transformations on the configuration space: The kinetic energy
operator (the Laplace operator) in the ordinary Schr\"odinger
equation provides a  
term {\em quadratic} in time derivatives in the 
path integral measure which is sufficient for the general
coordinate covariance \cite{lpi}. We remark that 
the regularization procedure based on the modified 
Schr\"odinger equation (\ref{ofwm}) applies to far more general
Hamiltonians than those quadratic in canonical momenta and
leading to the conventional {\em Lagrangian} path integral. 

The regularization procedure developed above deserves a further
mathematical elaboration. To this end we point out that the 
operator $\hat{K}_t^\nu$ acts in a Hilbert space, denoted
here and belows as ${\cal H}_\nu$, that is larger than the
original Hilbert space ${\cal H}_\infty$. Indeed, consider
canonical pairs $[\hat{Q}^k,\hat{P}_n]=i\delta_n^k,\,
[\hat{P}_k,\hat{P}_n]=[\hat{Q}^k,\hat{Q}^n]=0$ and sharp-eigenvalue
states $|q)$ (observe the difference in notation!) 
being eigenstates of the complete set of {\em
commutative} operators: $\hat{Q}^k|q)=q^k|q)$. These states
form a basis in ${\cal H}_\nu$. So a resolution of unity reads
$\hat{\1}_\nu = \int dq |q)(q|$ where $(q|q')= \delta(q-q')$.
In constrast to $|q)$, the coherent states $|q\>$ form a basis in 
the {\em smaller} space ${\cal H}_\infty$, though they carry the
same label $q$. Note that the coherent states in ${\cal H}_\nu$
would have twice as many labels, say $p$ and $q$, associated with
the canonical operators $\hat{P}$ and $\hat{Q}$, respectively.
In ${\cal H}_\nu$ the original canonical operators can be 
represented as $\hat{q}^k = -\oo^{kn}\hat{P}_n + \frac{1}{2}\hat{Q}^k$,
that is, we have a {\em reducible} representation of the original
Heisenberg algebra. It is also important to observe that the extension
of any opearator in ${\cal H}_\infty$ to the larger space ${\cal H}_\nu$
is by no means unique. This issue is not, however, our main concern.

To achieve a continuous-time regularization of the path integral 
measure, and thereby to make the latter mathematically well defined,
it is sufficient to consider a {\em specific} extension of the 
unitary operator $\hat{K}_t = \exp(-it\hat{H})$ which is induced
by the extension of the Hamiltonian (cf. (\ref{ofwm}))
\begin{equation}
\hat{H}\rightarrow \hat{H}_\nu = -\frac{i\nu}{2}\left(\hat{q}-
\hat{Q}\right)^2 + h(\hat{Q})\ ,
\end{equation}
where the function $h$ is the (lower) symbol of the original
Hamiltonian $\hat{H}$ in the canonical coherent state representation.
Observe that $h(\hat{Q})$ is uniquely defined, no operator ordering
ambiguities occur since the operators $\hat{Q}^k$ are commutative.
In the operator formalism developed, the Wiener measure regularization
is based on the important relation
\begin{equation}
\lim_{\nu\rightarrow\infty} (q|e^{-it\hat{H}_\nu}|q')\equiv
\lim_{\nu\rightarrow\infty} K_t^\nu(q,q') = \<q|e^{-it\hat{H}}|q'\>\ , 
\ \ t>0\ .
\label{ad1}
\end{equation}  
Thus, the ill-defined formal path integral representation of the r.-h.s.
is replaced by the {\em convergent} sequence of well-defined 
stochastic integrals in the l.-h.s. which, as has been argued above,
are explicitely covariant under general coordinate transformations
$q^k \rightarrow \bar{q}^k(q)$. Note also that $\hat{H}_\nu^\dagger 
\neq \hat{H}_\nu$, and the immaginary part of the extended 
Hamiltonian $\hat{H}_\nu$ gives rise to the Wiener measure 
in the path integral representation. 

Now let $q^i$ be local coordinates on a symplectic manifold, with
\begin{equation}
\{q^j,q^k\}_q = \omega^{jk}(q)\ .
\label{ssqq}
\end{equation}
The symplectic two-form $\omega = \omega_{ij}dq^i\wedge dq^j,
\omega_{ik}\omega^{kj}= \delta_i^j$, is closed $d\omega =0$
(the Jacobi identity for the Poisson bracket (\ref{ssqq})).
A formal path integral quantization (\ref{eopk}) (where
the symplectic volume element and the symplectic one-form
are modified according to the symplectic structure (\ref{ssqq}))
runs into notorious difficulties if one tries to relate it to
the operator formalism. In particular, by means of the formal change of
the integration variables to Darboux variables, the overlap
function $\<q|q'\>$
(the integral kernel of the unit operator, $h=0$)
seems always possible to be expressed as the integral (\ref{eopk})
with the canonical symplectic structure. This is certainly not true.
Coherent states on general manifolds do not coincide with the canonical 
coherent states. Therefore, the path integral measure must be 
regularized relative to general coordinate transformations on the 
phase space. A possible way to achieve this is to replace the 
formal, ill-defined Liouville measure by the Wiener measure
with an appropriate metric. However, there is a great deal
of freedom in choosing such a metric \cite{kl88}. The overlap function
$\<q|q'\>$ and, hence, the entire Hilbert space will depend on this
choice.  

The main goal of our work is to give a procedure of how the metric
can be constructed and to show that the corresponding path integral
is coordinate-free and determines a consistent quantum mechanics.

{\bf 2}. To avoid the difficult problem of quantizing
general symplectic structure (\ref{ssqq}), we shall use the conversion 
formalism \cite{c0,c}. The idea is to embed the symplectic manifold into a
larger Euclidean (flat) phase space and impose special second-class
constraints such that, when reduced on the constraint surface, 
the canonical symplectic structure 
of the embedding space turns into the original symplectic structure.
The second-class constraint system is then replaced by an effective 
Abelian gauge theory, and the latter is quantized by the standard
methods. The algebra of gauge invariant operators is then taken
as a quantum algebra of observables 
associated with the symplectic structure (\ref{ssqq}). 
The representation (Hilbert) space of this algebra is
a subspace of gauge invariant functions in the effective
quantum gauge theory.

Let us turn to the details. Consider an extended Euclidean phase space 
spanned by the Cartesian coordinates $q^i$ and momenta $p_i$, and
equipped with a canonical symplectic structure which we
choose as $\{q^i,q^j\}=\{p_i,p_j\}=0$ and $\{q^i,p_j\}=\delta_j^i$.
The second-class constraints are
\begin{equation}
\varphi_j = p_j + \alpha_j(q)=0\ , 
\label{sccpq}
\end{equation}
where $\alpha_i(q)$ are coefficients of the symplectic one-form
$\alpha = \alpha_idq^i$ and $d\alpha =\omega$ being the symplectic
two-form.
It is not hard to verify that $\Delta_{ij} = \{\varphi_i,\varphi_j\}=
\omega_{ij}(q)$. The symplectic structure on the constraint surface
is induced by the Dirac bracket $\{\cdot,\cdot\}_D = 
\{\cdot,\cdot\} - \{\cdot,\varphi_i\}\Delta^{ij}\{\varphi_j,\cdot\}$,
where $\Delta^{ik}\Delta_{kj}=\delta_j^i$. Eliminating the variables
$p_i$  by means of the constraints and assuming the variables $q^i$
to be the coordinates on the physical phase space, we find 
that the induced symplectic structure coincides with the original
one: $
\{q^j,q^k\}_D\vert_{\varphi =0} = \omega^{jk}(q)$.
Thus, a Hamiltonian dynamics on a generic symplectic
manifold can be described as a second-class constraint
system with the action
$S_\varphi = \int\left(p_j{\dot q}^j - h(q) 
- \lambda^j\varphi_j\right)dt$ on an extended phase space.
   
To quantize the theory, the canonical
variables $p_j$ and $q^j$ are promoted into self-adjoint operators
satisfying the canonical commutation relations ($[\,,\,] = i\{\,,\,\}$).
One next introduces a new set of canonical variables $\theta^j$
associated with the second-class constraints $\varphi_j$
also satisfying the Heisenberg commutation relations
$[\hat{\theta}{}^j,\hat{\theta}{}^k]=i\oo^{jk}$
and commuting with $\hat{q}{}^j$ and $\hat{p}{}^k$.
A general operator $\hat{A}$ can be represented in the form
\begin{equation}
\hat{A} = \sum_n A^{(n)}_{j_1\cdots j_n}(\hat{p},\hat{q}){\rm sym}
\left(\hat{\theta}{}^{j_1}\cdots\hat{\theta}{}^{j_n}\right)\equiv
 \sum_n A^{(n)}_{j_1\cdots j_n}(\hat{p},\hat{q}) \hat{e}_{(n)}^{j_1
\cdots j_n}\ , 
\label{symbasis}
\end{equation}
where the symbol sym means a symmetrized product of the operators
in the brackets. The physical subspace in the extended Hilbert space
is selected by the gauge invariance condition 
\begin{equation}
\hat{\sigma}_j |\psi_{ph}\> =0\ ,
\label{phs}
\end{equation}
where the generators of gauge transformations $\hat{\sigma}_j$ and
the Hamiltonian in the extended Hilbert space are constructed
to satisfy the conversion equations \cite{c0,c}
\begin{equation}
[\hat{\sigma}_j,\hat{\sigma}_k] =0 \ ,\ \ \ \
[\hat{H}_g,\hat{\sigma}_j]=0\ ,
\label{ce1}
\end{equation}
subject to the boundary conditions $
\hat{\sigma}_j^{(0)} = \hat{\varphi}_j$, $\hat{H}_g^{(0)} =
H(\hat{q})$. In the classical limit, the dynamics of the physical
degrees of freedom of the so constructed gauge theory coincides
with that on the original symplectic manifold \cite{c}, that is,
the correspondence principle is fulfilled. 

To solve the conversion equations, one sets \cite{bt,fl}
$\hat{\sigma}_j = \hat{\varphi}_j + A_j(\hat{q},\hat{\theta})$, $ 
A_j^{(0)}(\hat{q}) =0$.
The operators $\hat{q}^j$ commute, so we omit the hat while analyzing
the equation for $\hat{A}_j = A_j(q,\hat{\theta})$ that results from 
(\ref{ce1})
\begin{equation}
\pll_j \hat{A}_k -\pll_k\hat{A}_j - i[\hat{A}_j,\hat{A}_k]
=\omega_{jk}(q)\ .
\label{ym}
\end{equation}
Here $\pll_j = \pll/\pll q^j$. 
The solution is not unique. If $\hat{A}_k$ is a solution then 
for any $
\hat{\Lambda} = \Lambda(q,\hat{\theta})$, the operator
$\hat{A}_k^\Lambda = e^{i\hat{\Lambda}} \hat{A}_k e^{-i\hat{\Lambda}}
-i e^{i\hat{\Lambda}}\pll_k e^{-i\hat{\Lambda}}$
is also a solution, provided  this transformation does not
violate the zero boundary condition $\hat{A}_k^{(0)} =0$. In fact, 
this condition imposes some restrictions on $\hat{\Lambda}$ discussed later.

An important point to observe is that the physical amplitudes 
do not depend on $\hat{\Lambda}$. Indeed, any physical
amplitude reads $\<\psi_1|\hat{O}|\psi_2\>$ where $|\psi_{1,2}\>$
are gauge invariant (physical) states determined by (\ref{phs})
and $\hat{O}$ is any operator satisfying the condition 
$[\hat{O},\hat{\sigma}_i]\sim \hat{\sigma}_i$. If $\hat{A}_k$
is replaced by $\hat{A}_k^\Lambda$, then the new Abelian constraints
are $\hat{\sigma}_k^\Lambda = e^{i\hat{\Lambda}}\hat{\sigma}_k
e^{-i\hat{\Lambda}}$. Thus, the new sets of physical operators
and states are obtained from the former ones by the unitary
transformation $e^{i\hat{\Lambda}}$, that is, the physical amplitudes
are not changed. It is therefore sufficient to solve the conversion
equations modulo the  $\Lambda$-transformations. 

{\bf 3}. The right-hand side of Eq. (\ref{ym}) can be regarded as the
field strength in the Yang-Mills theory with an infinite
dimensional gauge group generated by the operators $\hat{e}_{(n)}$.
Introducing the Yang-Mills potential $\hat{C}_k = \alpha_k +
\hat{A}_k$, we observe that the corresponding field strength
vanishes if $\hat{A}_k$ satisfies (\ref{ym}). This means that
the vector potential $\hat{C}_k$ must be a pure gauge,
$\hat{C}_k = -i e^{i\hat{\Omega}}\pll_k e^{-i\hat{\Omega}}$,
where $\hat{\Omega}= \Omega(q,\hat{\theta})$. Yet, this
solution is determined modulo the gauge ($\Lambda$-)transformations.
The operator $\hat{\Omega}$ satisfies only the condition that
$\hat{C}_k^{(0)}=\alpha_k$ and this condition should not be violated
under the gauge transformations of $\hat{C}_k$. For this reason,
the pure gauge potential $\hat{C}_k$ can not be gauged to a vanishing
potential (which does not satisfy the condition  
$\hat{C}_k^{(0)}=\alpha_k$).

The operators $\hat{e}_{(0)}$ and $\hat{e}_{(1)}$ are generators
of the Heisenberg algebra which is a finite dimensional subalgebra
of the algebra generated by all the $\hat{e}_{(n)}$. So we can always
represent the gauge group element $e^{i\hat{\Lambda}}$ as the 
product  $e^{i\hat{\Lambda}_H}e^{i\hat{\Lambda}_\infty}$, where
$\hat{\Lambda}_H$ is an element of the Heisenberg algebra, and
$\hat{\Lambda}_\infty = \sum_{n>1} \lambda^{(n)}\hat{e}_{(n)}$
(here we suppress the indices $j_1,...,j_n$). The gauge transformation
of $\hat{C}_k$ with the gauge group element $e^{i\hat{\Lambda}_\infty}$
leaves the component $\hat{C}_k^{(0)}$ untouched so it does not
violate the required condition $\hat{C}_k^{(0)}=\alpha_k$. Thus,
modulo gauge transformations, we may write $\hat{C}_k = -i
e^{i\hat{\Omega}_H}\pll_k e^{-i\hat{\Omega}_H}$, where 
$\hat{\Omega}_H$ is an element from the Heisenberg algebra, or,
equivalently, 
\begin{equation}
\hat{C}_k = \alpha_k + \hat{A}_k =\alpha_k(q) +
a_{kj}(q)\hat{\theta}^j\ ,\ \ \ \ 
a_{kj}\oo^{ji}a_{in} = \omega_{kn}\ , \ \ \pll_ja_{ki}=\pll_ka_{ji}\ . 
\label{dg}
\end{equation}
The two latter relations follow from Eq.(\ref{ym}). If $Q^j=Q^j(q)$ 
are Darboux coordinates for the symplectic structure in question,
then in local coordinates $a_{jk} = \pll_jQ^{n}\oo_{nk}$.  

The gauge arbitrariness of the solution (\ref{dg}) is not yet
exhausted. The gauge transformations from the Heisenberg group
($\hat{\Lambda}=\hat{\Lambda}_H$) still apply. Representing
the gauge group element as the product of operators
$e^{i\lambda^{(1)}_j\hat{\theta^j}}$ (no summation over $j$)
and the phase factor $e^{i\lambda^{(0)}}$ (associated with 
the unit operator in the Heisenberg algebra), we calculate 
the gauge transformation of the solution (\ref{dg}) explicitly.
Assuming some set of the Darboux coordinates to represent 
the functions $a_{jk}$, we get that $Q^j\rightarrow Q^j_\lambda=
Q^j +\oo^{jk}\lambda^{(1)}_k$ under the gauge transformation. 
The condition that $\hat{C}_k^{(0)}=\alpha_k$ should not be violated 
under any gauge transformation would yield that $\pll_k\lambda^{(0)}
= f_k(\lambda^{(1)}, Q)$ (we omit the details) which specifies
$\lambda^{(0)}$. 
The integrability condition for this equation leads to
\begin{equation}
\pll_i Q_\lambda^k\oo_{kl}\pll_jQ_\lambda^l = \omega_{ij}\ .
\label{ic}
\end{equation}
Thus, $Q^j_\lambda$ is just another set of Darboux variables.
This, of course, might be expected at the very beginning 
since the residual gauge transformations
do not change the structure of the solution (\ref{dg}). 

The gauge freedom in choosing the Darboux variables appears desirable. Whatever
set of Darboux variables we use in practical calculations of
the physical amplitudes, the result will not depend on it.
Changing from one set of Darboux coordinates to another implies
a corresponding modification of the gauge invariance condition.
The physical states and operators are modified so that the 
physical amplitudes remain unchanged.
This freedom also allows one to make the quantum theory globally
well-defined \cite{fl}. 

{\bf 4}. The invariance of the theory under canonical transformations
of the Darboux variables is crucial for the path integral formalism
because one has to assume a specific form of the constraints in order
to calculate the path integral. So we pick out some set of Darboux
coordinates $Q^j$. The Hamiltonian of the effective gauge theory
satisfying the conversion equation (\ref{ce1}) has the form
\begin{equation}
\hat{H}_g= H(\hat{\vartheta})\ ,\ \ \ \hat{\vartheta}^i ={\rm sym}\,
q^i(\hat{\theta} + Q(q))\ ,
\label{hg}
\end{equation}
where the functions $q^i = q^i(Q)$ are inverse to $Q^i(q)$, i.e.,
$q^i(Q(q))\equiv q^i$. We also assume some appropriate ordering
of the operators $\hat{\vartheta}^i$ to make the Hamiltonian
hermitian. Making use of the explicit form of the constraints,
it is easy to prove that $[\hat{\sigma}_j,\hat{\vartheta}^k]=0$.
The required boundary condition follows from the relation 
$\hat{\vartheta}^{(0)\,i} = q^i$. 

The commutation relations of all canonical variables in the 
effective Abelian gauge theory are the standard ones, so we
can construct the Wiener measure regularized path integral
for the transition amplitude on the extended phase space
along the lines discussed in Section 1. To simplify notation, 
we denote by $b^\alpha$ the set $q^i,p_i$ and $\theta^i$, and
will write $\{b^\alpha,b^\beta\}=\oo^{\alpha\beta}$. The amplitude
$K^\nu_t(b,b')$ satisfies Eq.(\ref{ofwm}), where $q^i$
are now replaced by $b^\alpha$ and $h(b)$ is the symbol
of the operator (\ref{hg}) in the canonical coherent state representation.
The integral kernel  $K^\nu_t(b,b')$ is not physical because 
it is not gauge invariant. To obtain the physical evolution
operator, we make use of the projection method \cite{ufn,kl97}. Consider
a unitary operator $\hat{U}(\xi)= e^{i\xi^j\hat{\sigma}_j}$
with $\xi^j$ being parameters.
For any physical state we have $\hat{U}(\xi)|\psi\>=|\psi\>$.
We define a projection operator $\hat{\cal P} = \int 
d\kappa(\xi) \hat{U}(\xi)$
where  the measure $d\kappa(\xi)$ is normalized to unity, 
$\int d\kappa(\xi) =1$. Then $\hat{\cal P}^\dagger = \hat{\cal P} $
and $\hat{\cal P}^2 =\hat{\cal P}$. In general, the zero eigenvalue
of the constraint operators may lie in the continuum spectrum
of some or all the constraints. In this case we take a sequence
of the rescaled projection operators $c_\delta \hat{\cal P}_\delta,
c_\delta >0,$ as $\delta \rightarrow 0$, where 
$\hat{\cal P}_\delta = \int d\kappa_\delta(\xi) \hat{U}(\xi)$,
and the measure is chosen so that $\hat{\cal P}_\delta $ projects
the Hilbert space into a subspace where $\sum \hat{\sigma}_i^2 \leq
\delta^2$ for $0<\delta <\!\!< 1$ \cite{kl97}.

So, our task is to develop a coordinate-free path integral representation
of the gauge invariant transition amplitude $\<q|\hat{\cal P}\hat{K}_t
\hat{\cal P}|q'\> = \<q|\hat{K}_t
\hat{\cal P}|q'\>$. By construction, $[\hat{H},
\hat{\sigma}_k]=0$. To this end we consider the Schr\"odinger-like
equation that depend on a set of general functions $\xi^k(t)$
\begin{eqnarray}
i\pll_t\hat{K}_t^\nu[\xi]&=&
\left\{-\frac{i\nu}{2}\left(\hat{b}-\hat{B}\right)^2 + h(\hat{B}) +
\xi^k(t)\sigma_k(\hat{B})\right\}\hat{K}_t^\nu[\xi]\label{ad2}\\
&\equiv&
\{\hat{H}_\nu + \xi^k(t)\sigma_k(\hat{B})\}\hat{K}_t^\nu[\xi]\ .
\nonumber
\end{eqnarray}
As before, we have extended the original Hilbert space to 
an irreducible representation space of the canonical
operators $[\hat{B}^\alpha,\hat{P}_\beta]=i\delta_\beta^\alpha,\,
[\hat{B}^\alpha,\hat{B}^\beta]=[\hat{P}_\alpha,
\hat{P}_\beta]=0$, while
the original canonical operators now have a reducible
representation $\hat{b}^\alpha = -\oo^{\alpha\beta}\hat{P}_\beta 
+ \frac{1}{2}
\hat{B}^\alpha$. The functions $h$ and $\sigma$ are the (lower)
symbols of the Hamiltonian $\hat{H}$ and constraints
$\hat{\sigma}_k$ in the original Hilbert space ${\cal H}_\infty$.
With this choice we can prove the following important
property of the solution to Eq. (\ref{ad2}) with the intitial
condition $\hat{K}_{t=0}^\nu[\xi] =\hat{\1}_\nu$
\begin{equation}
(b|\hat{K}_t^\nu[\xi]|b') = 
(b|Te^{-i\int_0^td\tau[\hat{H}_\nu + \xi^k(\tau)\sigma_k(\hat{B})]}|b')
\rightarrow \<b|Te^{-i\int_0^td\tau[\hat{H}+\xi^k(\tau)\hat{\sigma}_k]}
|b'\>
\label{ad3}
\end{equation}
as $\nu\rightarrow \infty$, for all $t>0$. Here $\hat{B}^k|b)=
b^k|b)$. Averaging Eq.(\ref{ad3}) over the functions $\xi^k$
we obtain the sought-for representation of the physical transition
amplitude
\begin{equation}
\<b|\hat{K}_t\hat{\cal P}|b'\> =\lim_{\nu\rightarrow\infty}
\int d\rho(\xi)\, (b|\hat{K}_t^\nu[\xi]|b')\ , 
\label{ad4}
\end{equation}
where $d\rho(\xi)$ stands for a normalized  
measure for functions $\xi^k(\tau)$ on the interval 
$\tau\in [0,t]$. 
 In particular we may take 
the white-noise measure $d\rho(\xi) =d\rho_\nu(\xi)\equiv
{\cal D}\xi e^{-\frac{1}{2\nu}\int
  \xi^2d\tau}$, $\nu\rightarrow\infty$,
or a non-pinned Wiener measure \cite{klsh2}.
As a point of fact, the only condition the measure should satisfy
is that it must provide at least one action of the projection
operator ${\cal P}$ within the time interval $[0,t]$.
By going over to the path integral representation
in the l.-h.s. of Eq. (\ref{ad4}), we derive our final result
\begin{equation}
K_t(b,b') =\lim_{\nu\rightarrow\infty} 
N_\nu(t)\int d\rho(\xi)\!
\int\limits_{b(0)=b'}^{b(t)=b}\! d\mu_W^\nu(b)
e^{i\int_0^t \left(\frac 12 b^\alpha\oo_{\alpha\beta}db^\beta- 
[h(b)+ \xi^k\sigma_k(b)]d\tau
\right)}\ .
\label{wmrpig}
\end{equation}

The path integral (\ref{wmrpig}) with the measure $
  d\rho(\xi)$ included has a genuine (finite and countably additive)
measure which is covariant under general coordinate transformations.
It provides the sought for path integral formalism for general
Hamiltonian systems, which was our main goal.

To relate the path integral (\ref{wmrpig}) to the original
phase space manifold, one can make a change of variables
(which, in this case, is a well-defined procedure!) $(q,p,\theta)
\rightarrow (q,\sigma, \vartheta)$ where $\vartheta^k=
q^k(\theta + Q(q))$ or, equivalently, $\theta^k = \int^\vartheta_q
d\bar{q}^na_{nl}\oo^{lk}=\int^\vartheta_q dQ^k$ 
for any integration contour. In this representation,
the canonical pairs $(q,\sigma)$ are nonphysical degrees of freedom
due to the translational gauge symmetry $q\rightarrow q +\delta\xi$.
The variables $\vartheta$ are gauge invariant and represent the
physical degrees of freedom, and their Poisson bracket gives
the original symplectic structure $\{\vartheta^i,\vartheta^j\}=
\omega^{ij}(\vartheta)$ and $h=h(\vartheta)$ by construction. 
If the integral over $q,\sigma$ and $\xi$
is done, the result will be given by the exponential of the 
classical action averaged
with the {\em induced} non-canonical, non-flat Wiener measure
on the original symplectic manifold. The induced measure
will specify the unit operator kernel $\<\vartheta|\vartheta'\>$
and generalized coherent-state representation of the non-canonical
symplectic structure. This technical problem will be studied elsewhere.

{\bf 5}. In conclusion, we would like to emphasize the following.
A representation of the quantum mechanical evolution as a {\em
sum over paths} should not depend on coordinates because the 
geometrical notion of a path on the symplectic manifold is 
coordinate independent (a path is specified by its tangent
vector). Thus, in order to determine a formulation of quantum mechanics
via the path integral on the symplectic manifold, or,
in other words, to quantize a general Hamiltonian system
on the manifold by the path integral method, 
it is fundamental that the corresponding 
{\em functional} measure is coordinate-free.
The latter, in our opinion, requires an (induced) {\em metric} on the 
symplectic manifold.
We have 
given a constructive way of how this regularization can be obtained.
The procedure developed here provides a solution to the long-standing 
problem of 
constructing a rigorous coordinate-free path integral formalism on symplectic
manifolds. The above idea can also be successfully applied to
the path integral quantization of gauge and constrained systems, in
general (in a flat phase
space) \cite{klsh1,klsh2}. The 
operator technique established in Section 4 can be used to greatly
simplify the Wiener measure regularization of the path integral
in gauge theories and second-class constrained systems proposed
in our earlier works \cite{klsh1,klsh2}. It is also allows
one to generalize the result of \cite{klsh1} to most general
(open) first-class constraint algebras.

Thus, the coordinate-free path integral
formalism based on the Wiener measure regularization can be regarded
as a universal, self-consistent, path-integral quantization technique.

{\bf Acknowledgment}

I.A. Batalin is thanked for useful comments and references on
the conversion method.
S.V.S. is grateful to L. Baulieu and H. Kleinert for kind
hospitality at the  University Paris VI and the Free University of Berlin,
and to CNRS (France) and DFG (Germany) for financial support.

\end{document}